\documentclass[12pt,a4paper]{article}
\usepackage[latin1]{inputenc}
\usepackage{amsmath}
\usepackage{placeins}
\usepackage{url}
\usepackage{natbib}
\usepackage{color,xcolor,setspace,geometry}
\setcitestyle{authoryear,open={(},close={)}}
\usepackage{cite}
\usepackage{soul}
\soulregister\citep7
\soulregister\cite7
\soulregister\citet7
\soulregister\citealp7
\soulregister\ref7
\usepackage{amsfonts}
\usepackage{arydshln}
\usepackage{tikz}
\usetikzlibrary{shapes}
\usetikzlibrary{plotmarks}
\usetikzlibrary{tikzmark,matrix,calc}
\usetikzlibrary{arrows.meta,calc,decorations.markings,math,arrows.meta}
\usepackage{amssymb}
\usepackage{multirow}
\usepackage{graphicx}
\begin{document}
\begin{spacing}{1.35}	

\title{A copula-based multivariate hidden Markov model for modelling momentum in football}

\author{
Marius \"Otting\thanks{Bielefeld University},
Roland Langrock$^*$, Antonello Maruotti\thanks{Libera Universita Maria Ss. Assunta} \thanks{University of Bergen}\\[1em] 
}

\date{}

\maketitle

\begin{abstract} 
We investigate the potential occurrence of change points -- commonly referred to as ``momentum shifts'' -- in the dynamics of football matches. For that purpose, we model minute-by-minute in-game statistics of Bundesliga matches using hidden Markov models (HMMs). To allow for within-state correlation of the variables, we formulate multivariate state-dependent distributions using copulas. For the Bundesliga data considered, we find that the fitted HMMs comprise states which can be interpreted as a team showing different levels of control over a match. Our modelling framework enables inference related to causes of momentum shifts and team tactics, which is of much interest to managers, bookmakers, and sports fans. 
\end{abstract}

\section{Introduction}
Vocabulary such as ``momentum'', ``momentum shift'', or related terms is commonly used to refer to change points in the dynamics of a match.
Usage of such terms is typically associated with situations during a match where an event --- such as a shot hitting the woodwork in an association football match --- seems to change the dynamics of the match, e.g.\ in a sense that a team which prior to the event had been pinned back in its own half suddenly seems to dominate the match. 
A prominent example is the 2005 Champions League final between Milan and Liverpool, where Liverpool was trailing by three goals after the first half, but fought back after half time and eventually won by penalty shootout. 

Despite the widespread belief in momentum shifts in sports, it is not always clear to what extent \textit{perceived} shifts in the momentum are genuine. From the literature on the ``hot hand'' --- i.e.\ research on serial correlation in human performances --- it is well-known that most people do not have a good intuition of randomness, and in particular tend to overinterpret streaks of success and failure (see, e.g., \citealp{thaler2009nudge}, p.\ 30--34, and \citealp{kahneman2011thinking}, p.\ 114--118). It is thus to be expected that many perceived momentum shifts are in fact cognitive illusions in the sense that the observed shift in a competition's dynamics is driven by chance only.

Momentum shifts have been investigated in qualitative psychological studies, e.g.\ by interviewing athletes, who reported momentum shifts during matches (see, e.g., \citealp{richardson1988game, jones2008psychological}).
Fuelled by the rapidly growing amount of freely available sports data, quantitative studies have investigated the drivers of ball possession in football \citep{lago2010ball}, the detection of main playing styles and tactics \citep{diquigiovanni2018analysis, gonccalves2017exploring} and the effects of momentum on risk-taking \citep{lehman2013momentum}. In some of the existing studies, e.g.\ in \citet{lehman2013momentum}, momentum is not investigated in a purely data-driven way, but rather pre-defined as winning several matches in a row.

In this contribution, we analyse potential momentum shifts within football matches.
Specifically, we investigate the potential occurrence of momentum shifts by analysing minute-by-minute bivariate summary statistics from the German Bundesliga using hidden Markov models (HMMs). The corresponding data is described in Section \ref{chap:data}. Within the HMMs, we consider copulas to allow for within-state dependence of the variables considered. The corresponding methodology is presented in Section \ref{chap:models}. Our results, which are presented in Section \ref{chap:results}, suggest states which can be tied to different levels of control in a match. In addition, we investigate the causes of momentum shifts, e.g.\ the current score of the match. This type of insight could be of great interest to managers, bookmakers and sports fans.

\section{Data}\label{chap:data}
We analyse minute-by-minute in-game statistics of Bundesliga matches, taken from \url{www.whoscored.com}, to investigate to what extent momentum shifts in a football match are genuine, and what kind of events lead to a shift. Since the quality and tactics differ between the teams, we do not pool data from multiple teams, but consider data from a single team. Throughout this paper, we consider data from Borussia Dortmund. In the Supplementary Material, we present the same analysis for Hannover 96.

As proxy measures for the current momentum within a football match, we consider the number of shots on goal and the number of ball touches, with both variables sampled on a minute-by-minute basis. For match $m$, $m=1, \ldots, 34$, this results in a bivariate time series $\{ \mathbf{y}_{mt} \}_{t = 1, 2,\ldots, T_m}$, with $\mathbf{y}_{mt} = (y_{mt1}, y_{mt2})$ the pair of variables observed at time $t$ (out of $T_m$ minutes played) during the match.

Due to injury times being added to the regular match length of 90 minutes, the lengths of the time series considered range from 91 to 100 minutes. The final data set then comprises 3214 bivariate observations from 34 matches of the season 2017/18. In addition, since the underlying dynamics of a match, from Borussia Dortmund's perspective, potentially depend on characteristics of the opponent (such as the strength of the squad) as well as events in the match (such as goals), the following four covariates are considered:
\begin{itemize}
    \item the market value of the opposing team (taken from \url{www.transfermarkt.com});
    \item the goal difference in the current score;
    \item a dummy variable indicating whether the match is played at home or away;
    \item the current minute of the match. 
\end{itemize}
The first covariate considered is a (crude) proxy for the quality of teams. Specifically, a team's market value is given by the sum of all players' market values at the beginning of the season, and thus does not vary between matches or within matches, e.g.\ if players are substituted. The difference in the current score is calculated from Borussia Dortmund's point of view, i.e.\ positive values refer to a lead of Dortmund whereas negative values represent that Dortmund is trailing. The dummy indicating whether the match is played at home is included since several studies provided evidence for a home field advantage, because of (e.g.) crowd effects and psychological advantage when playing at home (see, e.g., \citealp{pollard2008home}). Finally, to account for the potential state of exhaustion of players, the minute of the match is also included. The variables considered are summarised in Table \ref{tab:descriptives}.

\begin{table}[!htbp] \centering 
  \caption{Descriptive statistics of the variables analysed, 'shots' and 'ball touches', as well as the covariates 'market value' and 'score difference'.} \vspace{0.5em}
  \label{tab:descriptives} 
\begin{tabular}{@{\extracolsep{5pt}}lccccc} 
\\[-1.8ex]\hline 
\hline \\[-1.8ex] 
 &  \multicolumn{1}{c}{Mean} & \multicolumn{1}{c}{St.\ Dev.} & \multicolumn{1}{c}{Min.} & \multicolumn{1}{c}{Max.} \\ 
\hline \\[-1.8ex] 
shots &  0.150 & 0.412 & 0 & 3 \\ 
ball touches & 6.101 & 5.036 & 0 & 28 \\ 
market value (in $10^6$ Euro) &  142.6 & 127.1 & 48.80 & 610.3 \\ 
score difference & 0.253 & 1.500 & $-$6 & 5 \\ 
\hline \\[-1.8ex] 
\end{tabular} 
\end{table}

\begin{figure}[!htb]
\centering
\includegraphics[scale=0.6]{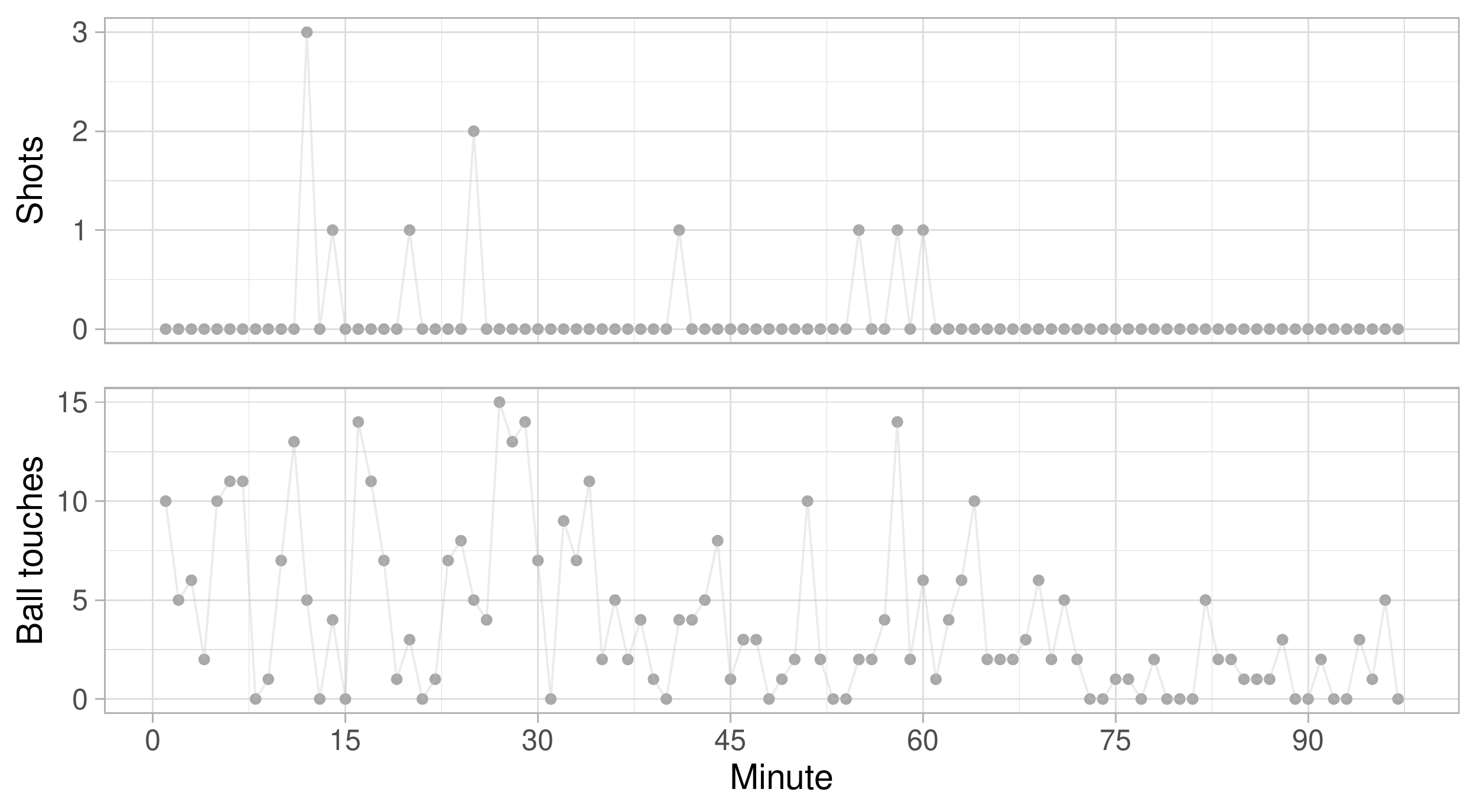}
\caption{Bivariate time series of the number of shots on goal (top) and the ball touches (bottom) of Borussia Dortmund for one example match from the data set (Borussia Dortmund vs.\ FC Schalke 04).} 
\label{fig:data}
\end{figure}

One example bivariate time series from the data set, corresponding to the in-game statistics observed for Borussia Dortmund in the match against FC Schalke 04 played in November 2017 is shown in Figure \ref{fig:data}. In the media, this match was said to have a momentum shift, since Borussia Dortmund was in a 4:0 lead at half time, but Schalke 04 scored four goals in the second half so that the match resulted in a draw.

\section{Modelling momentum}\label{chap:models}

Figure \ref{fig:data} underlines that there are periods in the match where Borussia Dortmund's number of ball touches and the number of shots on goal are fairly low (e.g.\ around minute 75--90), as well as periods with relatively many ball touches and shots on goal (e.g.\ around minute 15--30). HMMs hence constitute a natural modelling approach for the minute-by-minute bivariate time series data, as they accommodate the idea of a match progressing through different phases, with potentially changing momentum. The states can be interpreted as the underlying momentum, i.e.\ as potentially different levels of control of the team considered. 
In the simplest model formulation with two states, the states could, for example, be interpreted as either the team considered or the opponent having a high level of control (i.e.\ dominating the match). 
In this section, the basic HMM model formulation will be introduced (Section \ref{chap:baselinemodel}) and extended to allow for within-state dependence using copulas (Section \ref{chap:copulas}). The latter is desirable since the potential within-state dependence may lead to a more comprehensive interpretation of the states regarding the underlying momentum.
Finally, for the model formulation presented in Section \ref{chap:copulas}, covariates will be included (Section \ref{chap:covariates}).

\subsection{A baseline model}\label{chap:baselinemodel}
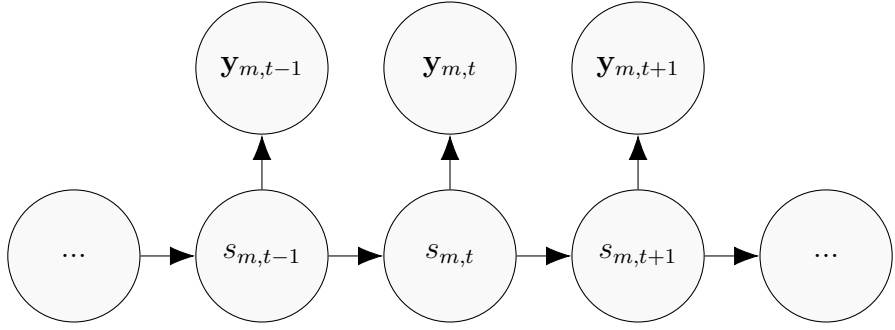
\begin{figure}
    \centering
	\begin{tikzpicture}
	\node[circle,draw=black, fill=gray!5, inner sep=0pt, minimum size=50pt] (A) at (2, -5) {$s_{m,t-1}$};
	\node[circle,draw=black, fill=gray!5, inner sep=0pt, minimum size=50pt] (A1) at (-0.5, -5) {...};
	\node[circle,draw=black, fill=gray!5, inner sep=0pt, minimum size=50pt] (B) at (4.5, -5) {$s_{m,t}$};
	\node[circle,draw=black, fill=gray!5, inner sep=0pt, minimum size=50pt] (C) at (7, -5) {$s_{m,t+1}$};
	\node[circle,draw=black, fill=gray!5, inner sep=0pt, minimum size=50pt] (C1) at (9.5, -5) {...};
	\node[circle,draw=black, fill=gray!5, inner sep=0pt, minimum size=50pt] (Y1) at (2, -2.5) {$\mathbf{y}_{m,t-1}$};
	\node[circle,draw=black, fill=gray!5, inner sep=0pt, minimum size=50pt] (Y2) at (4.5, -2.5) {$\mathbf{y}_{m,t}$};
	\node[circle,draw=black, fill=gray!5, inner sep=0pt, minimum size=50pt] (Y3) at (7, -2.5) {$\mathbf{y}_{m,t+1}$};
	\draw[-{Latex[scale=2]}] (A)--(B);
	\draw[-{Latex[scale=2]}] (B)--(C);
	\draw[-{Latex[scale=2]}] (A1)--(A);
	\draw[-{Latex[scale=2]}] (C)--(C1);
	\draw[-{Latex[scale=2]}] (A)--(Y1);
	\draw[-{Latex[scale=2]}] (B)--(Y2);
	\draw[-{Latex[scale=2]}] (C)--(Y3);
	\end{tikzpicture}
\caption{Dependence structure of the HMM considered: each pair of observations $\mathbf{y}_{mt}$ is assumed to be generated by one of $N$ (bivariate) distributions according to the state process $s_{mt}$.}
\label{fig:HMM}
\end{figure}

HMMs involve two components: an unobserved Markov chain with $N$ possible states, and an observed state-dependent process, whose observations are assumed to be generated by one of $N$ distributions as selected by the Markov chain. For the data considered in this paper, the observations and the state process are denoted by $\mathbf{y}_{mt}$ and $\{ s_{mt} \}_{t = 1,2,\ldots,T_m}$, respectively. Switches between the state are modelled by the transition probability matrix (t.p.m.) $\boldsymbol{\Gamma} = (\gamma_{ij})$, where $\gamma_{ij} = \Pr(s_{mt} = j | s_{m,t-1} = i),\, i,j = 1,\ldots,N$.
Figure \ref{fig:HMM} shows the model structure as directed graph.
For the model formulation of an HMM to be completed, the number of states $N$ and the class(es) of state-dependent distribution(s) have to be selected (see \citealp{zucchini2016hidden}, p.\ 29--31). While choosing state-dependent distribution(s) is straightforward for univariate time series, it is generally not straightforward to define a multivariate distribution to allow for within-state dependence of the variables considered. This would be straightforward if the marginals are assumed to be normally distributed, as in that case a multivariate normal state-dependent distribution can be used \citep[see, e.g.,][]{phillips2015objective}. However, as this assumption would here clearly be inadequate given that we consider count data,
for the vector of observations $\mathbf{y}_{mt}$ in the baseline model formulation we assume that the joint probability is obtained by the product of the marginal distributions,

\begin{equation}\label{eq:contmpcondind}
   f(\mathbf{y}_{mt} | s_{mt}) = \prod_{k=1}^K f(y_{mtk} \, |\, s_{mt}), 
\end{equation}
with $K=2$ here. This assumption, also known as \textit{contemporaneous conditional independence}, is often used in practice (see, e.g., \citealp{wall2009multiple, deruiter2017multivariate, punzo2018multivariate, van2019classifying}). Taking the product of the marginal distributions is thus straightforward and allows a flexible choice of the marginals $f(y_{mtk} \, |\, s_{mt})$, $k=1,\ldots,K$. In Eq.\ (\ref{eq:contmpcondind}), each of these denotes a probability mass function (p.m.f.) since we deal with discrete data, but in principle \textit{f} could also denote a density without any further changes in the baseline model formulation. The $K=2$ variables modelled here will still be unconditionally dependent when assuming contemporaneous conditional independence, as the underlying Markov chain induces both serial dependence and cross-dependence between them.
The contemporaneous conditional independence assumption will be modified in the next subsection.

Since both the number of shots on goal and the number of ball touches are count data, the Poisson distribution would be a standard choice for either of the two variables. Here, to account for possible over- and underdispersion in the data, a Conway-Maxwell-Poisson (CMP) distribution is assumed both for the number of shots on goal and the number of ball touches, with p.m.f.\

$$
\Pr(X = x) = \dfrac{1}{Z(\lambda, \nu)} \dfrac{\lambda^x}{(x!)^\nu},
$$
with $Z(\lambda, \nu) = \sum_{k = 0}^{\infty} \lambda^k / (k!)^\nu$, $\lambda > 0$ and $\nu \ge 0$ \citep{cmp}. The CMP distribution contains some well-known discrete distributions:
\begin{itemize}
\item for $\nu=1$,  $Z(\lambda,\nu)=e^{\lambda}$, and the CMP distribution simply reduces to the ordinary Poisson($\lambda$);
\item for $\nu\rightarrow\infty$,  $Z(\lambda,\nu)\rightarrow 1+\lambda$, and the CMP distribution approaches the Bernoulli with parameter  $\lambda(1+\lambda)^{-1}$;
\item for $\nu=0$ and $0<\lambda<1$, $Z(\lambda,\nu)$ is a geometric sum $$Z(\lambda,\nu) =\sum_{j=0}^{\infty}\lambda^j= \frac{1}{1-\lambda}$$ and, accordingly, the CMP distribution reduces to the geometric distribution $p_x = \lambda^x(1-\lambda)$;
\item  for $\nu = 0$ and $\lambda\geq 1$, $Z(\lambda,\nu)$ does not converge, leading to an undefined distribution.
\end{itemize}
In general, the normalising constant $Z(\lambda,\nu)$ does not reduce to such a simple closed-form expression. Asymptotic results are however available \citep{gillispie2015approximating}.

To formulate the likelihood for the baseline model, the $i-$th diagonal element of the $N \times N$ diagonal matrix $\mathbf{P}(\mathbf{y}_{mt})$ consists of the joint probability of the observations $y_{mt1} \text{ and } y_{mt2}$ given state $i$, i.e.\ $f(y_{mt1} \, |\, s_{mt} = i) \cdot f(y_{mt2} \, |\, s_{mt} = i)$. Since the Conway-Maxwell-Poisson distribution contains an infinite sum in the normalising constant, the evaluation of the p.m.f.\ is not straightforward. Here, the R package \texttt{COMPoissonReg} was used for this purpose \citep{compoisson}. 
Since stationarity cannot reasonably be assumed in our setting, we estimate the initial distribution $\boldsymbol{\delta}= \big(\Pr (s_{m1} = 1),\ldots,\Pr (s_{m1} = N) \big)$, regarding the parameters of  $\boldsymbol{\delta}$ as $N-1$ additional parameters to be estimated.
With these quantities defined, the likelihood for a single match $m$ is given by:
\begin{equation*}
L = \boldsymbol{\delta} \mathbf{P}(\mathbf{y}_{m1}) \mathbf{\Gamma}\mathbf{P}(\mathbf{y}_{m2}) \dots \mathbf{\Gamma}\mathbf{P}(\mathbf{y}_{mT_m}) \mathbf{1}
\end{equation*}
with column vector $\mathbf{1}=(1,\ldots,1)' \in \mathbb{R}^N$ (see \citealp{zucchini2016hidden}, p.\ 37). Calculation of this matrix product expression amounts to the application of the forward algorithm, which is a powerful recursive technique for efficiently calculating the likelihood of an HMM at computational cost $\mathcal{O}(TN^2)$ only (see \citealp{zucchini2016hidden}, p.\ 38). To obtain the likelihood for the full data set, we assume independence between the individual matches. The likelihood is thus given by the product of likelihoods for the individual matches:

\begin{equation}\label{eq:HMMlike}
L = \prod_{m=1}^{34} \boldsymbol{\delta} \mathbf{P}(\mathbf{y}_{m1}) \mathbf{\Gamma}\mathbf{P}(\mathbf{y}_{m2}) \dots \mathbf{\Gamma}\mathbf{P}(\mathbf{y}_{mT_m}) \mathbf{1}
\end{equation}
The model formulation presented here could be extended to account for momentum carry-over effects across matches, but this is not investigated in the present work since there is usually a time difference of 5-7 days between matches.
The model parameters are estimated by numerical maximum likelihood estimation using the function \texttt{nlm()} in R \citep{rcoreteam}. To avoid local maxima, we selected starting values for the numerical maximisation by drawing random numbers from uniform distributions 50 times and choosing the model with the best likelihood. An exploratory analysis guided the choice of what constitutes reasonable ranges for the parameter values for the state-dependent distributions.
For a model with $N=2$ states, it took less than a minute to numerically maximise the likelihood on a standard desktop computer. In the Supplementary Material of this article, we provide data and code for all models presented.

\subsection{Modelling within-state dependence using copulas}\label{chap:copulas}
In the baseline model formulation, we assume contemporaneous conditional independence, i.e.\ that there is no within-state correlation between the two variables considered. However, when modelling momentum in football, it is of interest to explicitly model any within-state dependence to draw a comprehensive picture of the dynamics of a match. For example, high ball possession can be linked to both an attacking phase with lots of shots on goal, but also much less goal-oriented tactics, where the main aim is simply to control the match by keeping possession of the ball, without much pressure on goal. The between-variable correlation would likely be very different in those two scenarios. By estimating the within-state correlation between the two variables, we are better able to distinguish between such fairly subtle differences in a team's style of play.

To modify the contemporaneous conditional independence assumption, a multivariate distribution needs to be assumed to specify the dependence structure between the variables considered within states.
Here, we allow for within-state correlation of our variables $\mathbf{y}_{mt}$ by formulating a bivariate distribution as state-dependent distribution using a copula. A copula is a multivariate probability distribution with uniform margins. As introduced by \citet{sklar1959fonctions}, the idea of a copula is to split a multivariate distribution into its univariate margins and the dependence structure, where the latter depends on the copula considered. 
Within the class of HMMs, copulas have previously been used by \citet{hardle2015hidden} to model within-state dependence in financial data, and by \citet{brunel2005unsupervised} and \citet{lanchantin2011unsupervised} for image analysis.
For our modelling approach, we again consider the Conway-Maxwell-Poisson both for the number of shots on goal and the number of ball touches as marginal distribution. With $F_1(y_{mt1} | s_{mt})$ and $F_2(y_{mt2} | s_{mt})$ denoting the (state-dependent) cumulative distribution function of the marginals, the bivariate state-dependent distribution is given by

$$
F(\mathbf{y}_{mt} \, | \, s_{mt}) = C\big(F_1(y_{mt1} \, | \, s_{mt}), F_2(y_{mt2}\, | \, s_{mt})\big),
$$
where $C(.,.)$ is a bivariate copula. When deriving the corresponding p.m.f., differences are needed rather than derivatives, since the marginals are discrete (see, e.g., \citealp{nikoloulopoulos2013copula}). Thus, the bivariate p.m.f.\  of $\mathbf{y}_{mt}$ given state $s_{mt}$ is given by

\begin{equation}\label{eq:discretecopula}
\begin{split}
f(\mathbf{y}_{mt}\, |\, s_{mt}) = & \: C\big(F_1(y_{mt1}\, |\, s_{mt}), F_2(y_{mt2}\, |\, s_{mt}) \big) \\ & - C\big(F_1(y_{mt1} - 1\, |\, s_{mt}), F_2(y_{mt2}\, |\, s_{mt})\big) \\
& - C\big(F_1(y_{mt1}\, |\, s_{mt}), F_2(y_{mt2} - 1\, |\, s_{mt})\big) \\
& + C\big(F_1(y_{mt1} - 1\, |\, s_{mt}), F_2(y_{mt2} - 1\, |\, s_{mt} )\big).
\end{split}
\end{equation}
The copula $C(.,.)$ needs to be selected from the large number of possible copula functions available in the literature. Here, we focus on copulas that can model positive and negative dependence. Archimedean copulas (see, e.g., \citealp{nelsen2007introduction}, p.\ 116--118, for an overview) are convenient for this modelling purpose. We consider three different families of copulas, comparing their fit to the data in Section \ref{chap:results}: first, the Frank copula, which for two marginals $u_1$ and $u_2$ defined as
$$
C(u_1, u_2) = -\dfrac{1}{\theta} \log\Big(1 + \dfrac{(\exp(-\theta u_1) - 1) (\exp(-\theta u_2) - 1)}{\exp(-\theta) - 1} \Big), \, \theta \in \mathbb{R} \setminus \{0\},
$$
second, the Clayton copula,
$$
C(u_1, u_2) = \Big(\max\{u_1^{-\theta} + u_2^{-\theta} - 1; 0 \}\Big)^{-1/\theta}, \, \theta \in [-1; \, \infty) \setminus \{0\},
$$
and third, the Ali-Mikhail-Haq (AMH) copula,
$$
C(u_1, u_2) = \dfrac{u_1 u_2}{1 - \theta (1 - u_1) (1 - u_2)}, \, \theta \in [-1, \, 1),
$$
where for each copula considered the dependence parameter is denoted by $\theta$.
As $\theta\rightarrow 0$, each of the three copulas above approaches the independence copula. For the Frank copula, as $\theta\rightarrow \infty$, the copula converges to the co-monotonicity copula corresponding to perfect positive dependence, while for $\theta\rightarrow -\infty$ it converges to the counter-monotonicity copula corresponding to perfect negative dependence.
For the Clayton copula, as $\theta \rightarrow -1$ ($\theta \rightarrow \infty$), the copula converges to the counter-monotonicity (co-monotonicity) copula with perfect dependence. The AMH copula converges to neither the co-monotonicity nor the counter-monotonicity copula (see \citealp{nelsen2007introduction}, p.\ 116--118).

With the copulas defined as above, the diagonal matrix $\mathbf{P}(\mathbf{y}_{mt})$ in the HMM likelihood (see Eq.\ \ref{eq:HMMlike}) changes slightly. The
$i$--th diagonal entry is now equal to $f(\mathbf{y}_{mt} | s_{mt} = i)$ as defined in Eq.\ (\ref{eq:discretecopula}) instead of the product of the marginals. The corresponding likelihood is then again numerically maximised using the function \texttt{nlm()} in R. For that purpose, we again carefully selected different starting values, as it was done for the baseline model introduced above.

\subsection{A model including covariates}\label{chap:covariates}
In the previous subsections, the transition probabilities $\gamma_{ij}$ were assumed to be constant over time. To account for possible events which may lead to state-switching, and hence to possible momentum shifts, we modify this assumption by explicitly allowing the transition probabilities $\gamma_{ij}$ to depend on covariates at time $t$. This is done by linking $\gamma_{ij}^{(t)}$ to covariates $x_1^{(t)},\ldots,x_p^{(t)}$ using the multinomial logit link:
$$
\gamma_{ij}^{(t)} = \dfrac{\exp(\eta_{ij}^{(t)})}{\sum_{k=1}^N \exp(\eta_{ik}^{(t)})}
$$
with \vspace{0.5cm}
$$ \eta_{ij}^{(t)} = 
\begin{cases}
\eta_{ij}^{(t)} = \beta_0^{(ij)} + \sum_{l=1}^p \beta_l^{(ij)} x_l^{(t)}  & \text{if }\, i\ne j; \\
0 & \text{otherwise}.
\end{cases} 
\vspace{1cm}
$$
Since the transition probabilities depend on covariates, the t.p.m.\ $\boldsymbol{\Gamma}_t$ is not constant across time anymore, i.e.\ the Markov chain is non-homogeneous. However, the structure of the HMM likelihood as stated in Eq.\ (\ref{eq:HMMlike}) is unaffected, i.e.\ the likelihood can still be maximised numerically, again with different sets of starting values to avoid local maxima.

\section{Results}\label{chap:results}
In this section, the different models presented in Section \ref{chap:models} are fitted to data on the matches of Borussia Dortmund in the 2017/18 Bundesliga season. To further illustrate the methodology, in particular for lower-ranked teams, in the Supplementary Material we provide the results also for Hannover 96.

\subsection*{\normalsize \textbf{\textit{Baseline model}}}

For the baseline model, we make use of the contemporaneous conditional independence assumption, cf.\ Eq.\ (\ref{eq:contmpcondind}), initially focusing on the case of $N=2$ states. The corresponding parameter estimates associated with the number of shots on goal are $\boldsymbol{\hat{\lambda}}_{\text{shots}} = (0.125, 0.149)$, $\boldsymbol{\hat{\nu}}_{\text{shots}} = (0.206, 0.001)$, while for the number of ball touches, they are $\boldsymbol{\hat{\lambda}}_{\text{touches}} = (0.971, 2.381)$, $\boldsymbol{\hat{\nu}}_{\text{touches}} = (0.102, 0.390)$. It is not straightforward here to compute the means of the fitted distributions due to the infinite sum in the normalising constant. \citet{macdonald2018time} discuss several approaches and calculate the mean by $\frac{1}{Z(\lambda, \nu)} \sum_{k=0}^d k \lambda^k / (k!)^\nu$ using a very large $d$ (say $d=100$). Following this approach, the means of the number of shots on goal are 0.138 and 0.175 for states 1 and 2. For the ball touches, the means are 4.080 (state 1) and 10.104 (state 2). Thus, state 2 can be interpreted as the team considered, Borussia Dortmund, being more dominant, i.e.\ having a higher level of control over the match, than when being in state 1. The t.p.m.\ is estimated as

\begin{align*} 
\hat{\mathbf{\Gamma}} = 
\begin{pmatrix}
0.867 & 0.133 \\
0.280 & 0.720  \\
\end{pmatrix},
\end{align*} 
and the initial distribution as $\boldsymbol{\hat{\delta}} = (0.258, 0.742)$. According to the t.p.m.\ of the fitted model, there is some persistence in both states. Although this is the simplest model formulation considered here, the fitted model comprises interpretable states which refer to different levels of control over the match. The model can thus be regarded as a simple baseline model for capturing momentum shifts. We will now gradually increase its complexity to more fully capture the in-game dynamics.

\subsection*{\normalsize \textbf{\textit{Copula-based HMM with}} $\boldsymbol{N}$ \textbf{\textit{= 2}}}

To capture possible within-state correlation of the variables, a multivariate distribution needs to be considered. For Poisson marginals, the bivariate Poisson as proposed by \citet{karlis2003analysis} would be a possible candidate. However, as discussed in Section \ref{chap:baselinemodel}, this approach would have two limitations, namely the inability to capture overdispersion (or underdispersion) in the observations, and the restriction to positive between-variable correlation. Instead we use more flexible CMP distributions for the marginals, stitching them together using a copula as described in Section \ref{chap:copulas}.

\begin{figure}[!htb]
\centering
\includegraphics[scale=0.7]{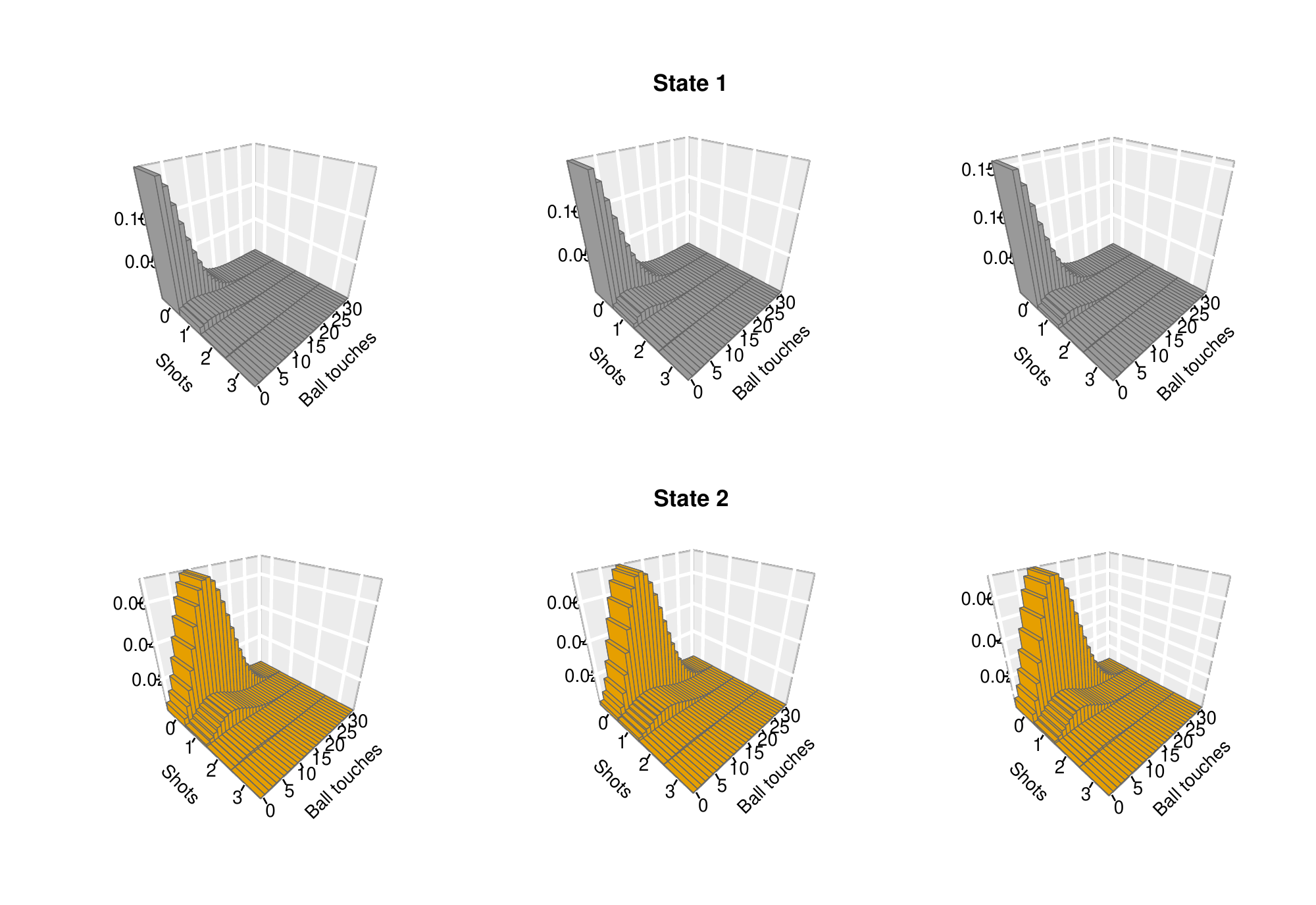}
\caption{Fitted state-dependent distributions for the baseline two-state HMM for Borussia Dortmund. From left to right: Frank-, Clayton-, and AMH-copula.} 
\label{fig:baselineDO}
\end{figure}

First, we investigate the consequences of relaxing the contemporaneous conditional independence assumption. To this end, Figure \ref{fig:baselineDO} displays the estimated state-dependent distributions of two-state copula-based HMM formulations, using the Frank, Clayton, and AMH copula. While visually there is no clear difference between the different copula functions considered, the application of the Clayton copula led to the highest likelihood of the fitted model. Compared to the baseline model, the copula-based model shows a clear improvement in the fit ($\Delta \text{AIC} = 48; \Delta \text{BIC} = 35$). The fitted state-dependent distributions can again be interpreted as Borussia Dortmund exhibiting different levels of control, with state 1 corresponding to situations where the game is balanced, whereas state 2 refers to a high level of control. As for the baseline model, there is a fairly high persistence in the states, with the diagonal elements of the t.p.m.\ estimated as $\hat{\gamma}_{11} = 0.852$ and $\hat{\gamma}_{22} = 0.706$.

\subsection*{\normalsize \textbf{\textit{Choosing the number of states}}}
For the choice of the number of states, it is anything but clear how many states a given team may exhibit in a football match. To choose an appropriate number of states, and also a copula, we first consult the AIC and the BIC for the copula-based HMMs using different numbers of states and the three copulas considered above. The corresponding results are displayed in Table \ref{tab:AICBIC}. Starting with the choice of the copula, the Clayton copula is preferred by both AIC and BIC. Hence, from now on, we use the Clayton copula. However, we note that when considering different marginal distributions, the fit of the copula also depends on the fit of the marginal distributions, which generally renders the choice of the copula a challenging task (see \citealp{mikosch2006}).
For the number of states, the choice is not as conclusive: according to the AIC, the five-state model is preferred, whereas the BIC selects three states. As it is well-known that the AIC tends to select too many states in a HMM (see \citealp{pohle2017selecting}), a choice of $N=3$ seems more appropriate based on these formal criteria. To make an informed choice based also on interpretability of the resulting model states, in Figure \ref{fig:clayton34states} we further inspect the fitted models with three and four states, by means of their estimated state-dependent distributions. Figure \ref{fig:clayton34states} illustrates that the general patterns of the state-dependent distributions from the three-state model are also included in the four-state model, whereas the state-dependent distribution of state 2 in the four-state model seems to refer to an underlying level of control which is not included in the three-state model. However, at closer inspection of the distributional shapes in the four-state model, there is a substantial overlap between the state-dependent distributions of state 2 and state 3. Hence, given that the BIC points to the three-state model, and since we do not see meaningful additional information in a potential fourth state, from now on we focus exclusively on three-state models.

\subsection*{\normalsize \textbf{\textit{Copula-based HMM with}} $\boldsymbol{N}$ \textbf{\textit{= 3}}}
\begin{figure}[!htb]
\centering
\includegraphics[scale=0.65]{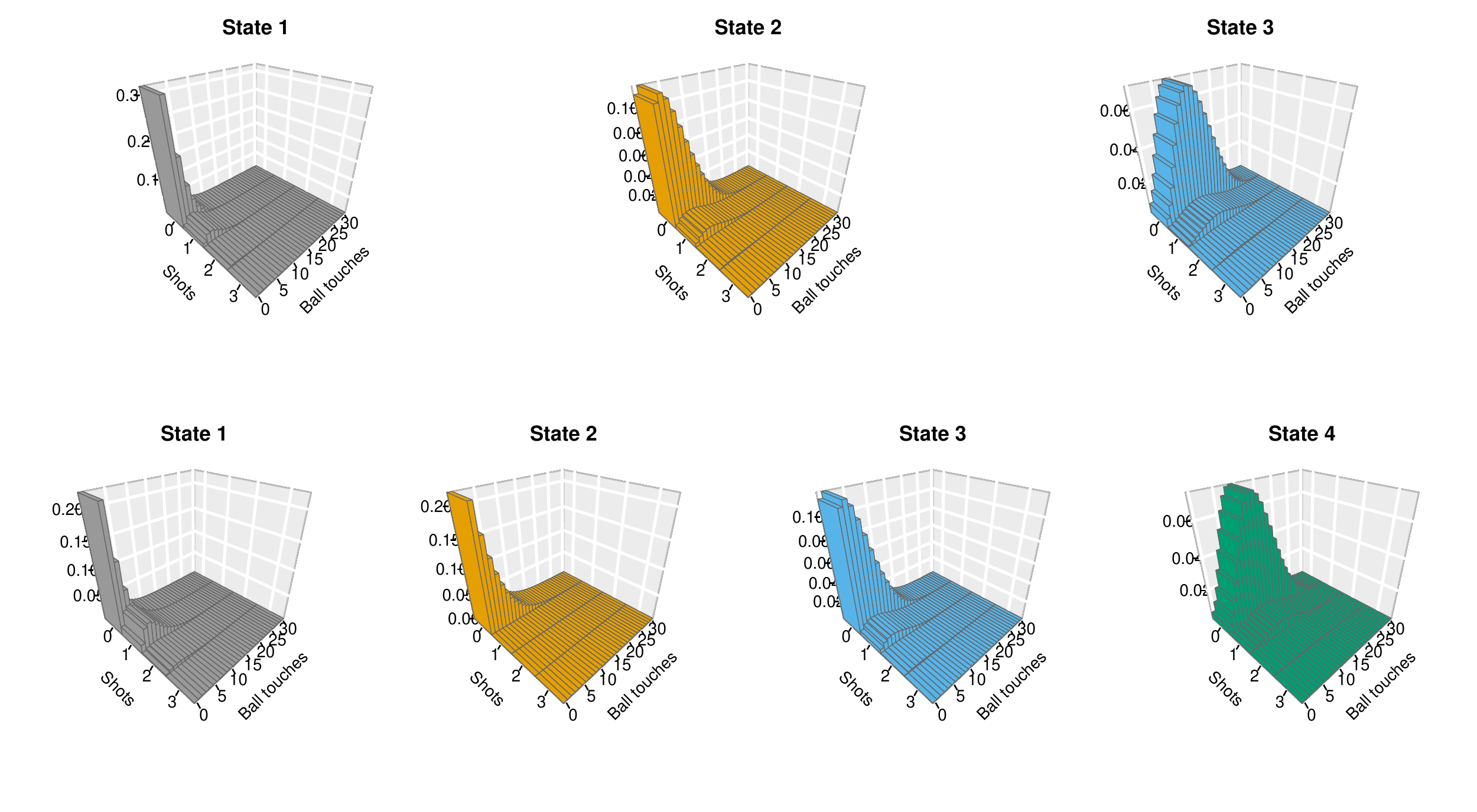}
\caption{State-dependent distributions for the three-state (top row) and four-state (bottom row) Clayton-copula HMM.} 
\label{fig:clayton34states}
\end{figure}

For the Clayton-copula HMM with three states, Table \ref{tab:3statenocov} displays the estimated parameters of the marginal distributions as well as the dependence parameter of the copula. Deriving the corresponding means for the marginal distributions as described above yields means for the number of shots of 0.226, 0.132, and 0.147 for state 1, 2, and 3. For the number of ball touches, the corresponding means are 2.032 (state 1), 4.583 (state 2), and 9.732 (state 3). Based on the means and the corresponding distributional shapes (see top row in Figure \ref{fig:clayton34states}), the different states can be interpreted as Borussia Dortmund showing different levels of control over the match: low control in state 1, a fairly balanced match in state 2, and high control with lots of ball possession in state 3. State 1, with its relatively high mean number of shots on goal despite the fewer ball touches, likely includes several different styles of play with a low level of control, e.g.\ a defensive style of play, counter attacks, and situations like (counter-)pressing. In state 3, the estimated negative dependence between the number of shots and ball touches may result from two different styles of high-control play: either Borussia Dortmund is controlling and passing the ball without much pressure on goal, or they go effectively straight for goal, without much passing. In addition, the t.p.m.\ is estimated as

\begin{align*} 
\hat{\mathbf{\Gamma}} = 
\begin{pmatrix}
0.471 & 0.054 & 0.475 \\
0.006 & 0.988 & 0.006  \\
0.195 & \approx 0 & 0.805
\end{pmatrix}.
\end{align*} 
Here, with $\hat{\gamma}_{22} = 0.988$ and $\hat{\gamma}_{33} = 0.805$, there is very high persistence in state 2 (balanced state) and moderately high persistence in state 3 (high-control state). Staying in state 1 (low control and quick counter attacks) is relatively unlikely ($\hat{\gamma}_{11} = 0.471$), and switching to the high-control state when being in state 1 is most likely. Up next we will present the results for the model including covariates in the state process.

\begin{table}
\caption{AIC and BIC for copula-based HMMs with different numbers of states.}\vspace{0.5em}
\label{tab:AICBIC}
\centering
\begin{tabular}{lll:ll:ll}
 & \multicolumn{2}{c}{Frank} & \multicolumn{2}{c}{Clayton} & \multicolumn{2}{c}{AMH} \\ 
 & AIC & BIC & AIC & BIC & AIC & BIC \\ \hline
2 states & 20,954 & 21,033 & 20,941 & 21,020 & 20,943 & 21,022 \\
3 states & 20,865 & \textbf{21,005} & 20,839 & \textbf{20,979} & 20,861 & \textbf{21,001} \\
4 states & 20,836 & 21,049 & 20,817 & 21,030 & \textbf{20,831} & 21,043 \\
5 states & \textbf{20,814} & 21,112 & \textbf{20,801} & 21,098 & 20,834 & 21,132 \\
\end{tabular}
\end{table}

\begin{table}
\caption{Parameter estimates for the state-dependent distributions of the Clayton-copula HMM with three states.}\vspace{0.5em}
	\centering
	\scalebox{0.9}{
	\begin{tabular}{cccc}
	Variable & State 1  & State 2  & State 3 \\[-2pt] \hline \\[-6pt]
	Shots on goal & $\hat{\lambda} = 0.212, \hat{\nu} = 0.631$ & $\hat{\lambda} = 0.117, \hat{\nu} \approx 0$ & $\hat{\lambda} = 0.128, \hat{\nu} = 0.002$ \\[5pt]  
	Ball touches & $\hat{\lambda} = 0.670, \hat{\nu} \approx 0$ & $\hat{\lambda} = 1.093, \hat{\nu} = 0.149$ & $\hat{\lambda} = 2.145, \hat{\nu} = 0.352$ \\[5pt]   
	Dependence & $\hat{\theta} = 1.721$ & $\hat{\theta} = 0.510$ & $\hat{\theta} = -0.048$ \\ 
	\hline 
    \end{tabular}}
\label{tab:3statenocov}
\end{table}

\subsection*{\normalsize \textbf{\textit{A model including covariates}}}
The models presented so far already provide interesting insights into the dynamics of football matches, since the state-dependent distributions can be tied to different levels of control of the team considered. To gain further insights, we incorporate covariates to investigate potential drivers of momentum shifts. According to the AIC, the model including all covariates considered is preferred over the model without covariates ($\Delta \text{AIC} = 51$); we do not conduct variable selection as we regard this analysis step as explanatory (rather than an attempt to find the best model).

For ease of interpretation, we visualise the estimated transition probabilities as functions of covariates, and present the theoretical stationary distributions of the Markov state process when fixing the covariate values at certain levels. The theoretical stationary distributions indicate how state occupancy, i.e.\ how much time is spent in a state, varies across different values of the covariate considered \citep{patterson2009classifying}. 
To illustrate these two approaches, we present (i) the transition probabilities as functions of the covariate minute, and (ii) the stationary distributions with respect to the score difference. In Table \ref{tab:appendix_coefficients} in the Supplementary Material, the estimated $\beta_0^{(ij)}, \ldots, \beta_p^{(ij)}$ and their 95\% CIs are displayed.

\begin{figure}[!htb]
\centering
\includegraphics[scale=0.8]{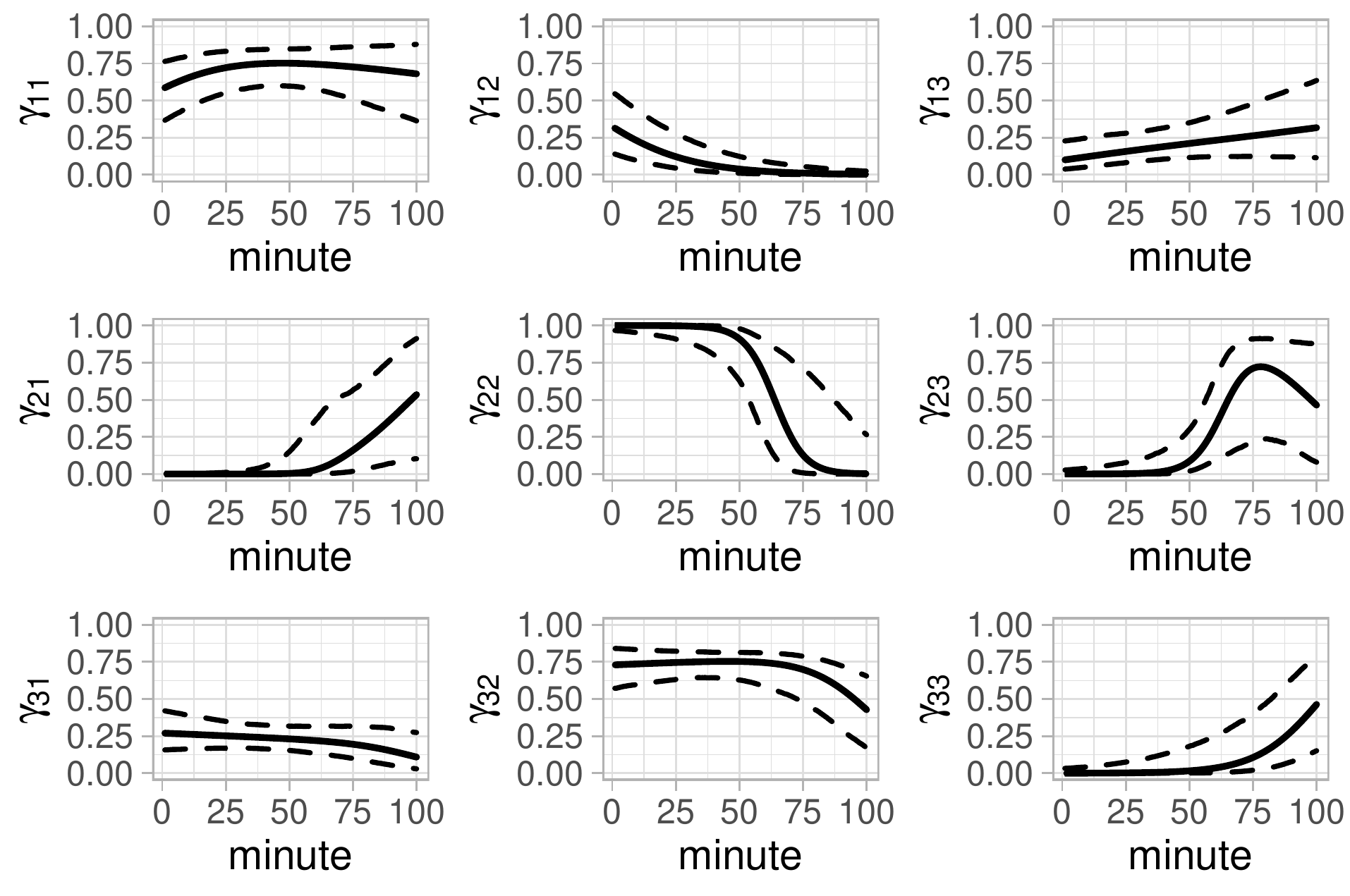}
\caption{Transition probabilities as functions of the covariate minute. The dashed lines indicate confidence intervals (obtained based on Monte Carlo simulation). The values of the score difference and the market value of the opponent are set to 0 and 200. Table \ref{tab:appendix_coefficients} in the Supplementary Material displays the coefficients of the multinomial logistic regression underlying this figure.}
\label{fig:coveffects}
\end{figure}

For (i), as displayed in Figure \ref{fig:coveffects}, the values of the score difference and the market value of the opponent are set to 0 and 200, corresponding to situations where the score is even and the opponent's strength is about average. In addition, we focus on home matches only, since the corresponding dummy variable in the linear predictor does not affect the overall pattern regarding the direction of the effect.
The confidence intervals (indicated by the dashed lines) are obtained based on Monte Carlo simulation from the approximate multivariate normal distribution of the estimator. 
According to the estimated effects, switching from state 1 (low control and quick counter attacks) and state 2 (balanced state) to state 3 (high-control state), becomes more likely at the end of matches.
In addition, staying in state 3 also becomes more likely at the end of matches.

The stationary distributions for the score difference are shown in Table \ref{tab:scorediff_stat}. The values of the minute and the market value of the opponent are fixed at 80 and 200, corresponding to situations in the final stage of a match with the opponent's strength being about average.
The stationary distributions indicate that there is a high probability for Borussia Dortmund to be in state 3 (high-control state) either if they have a clear lead or if they are trailing. In contrast, if they hold only a slender lead, then the probability of being in state 1 (low control and quick counter attacks) is highest. 

\begin{figure}[!htb]
\centering
\includegraphics[scale=0.62]{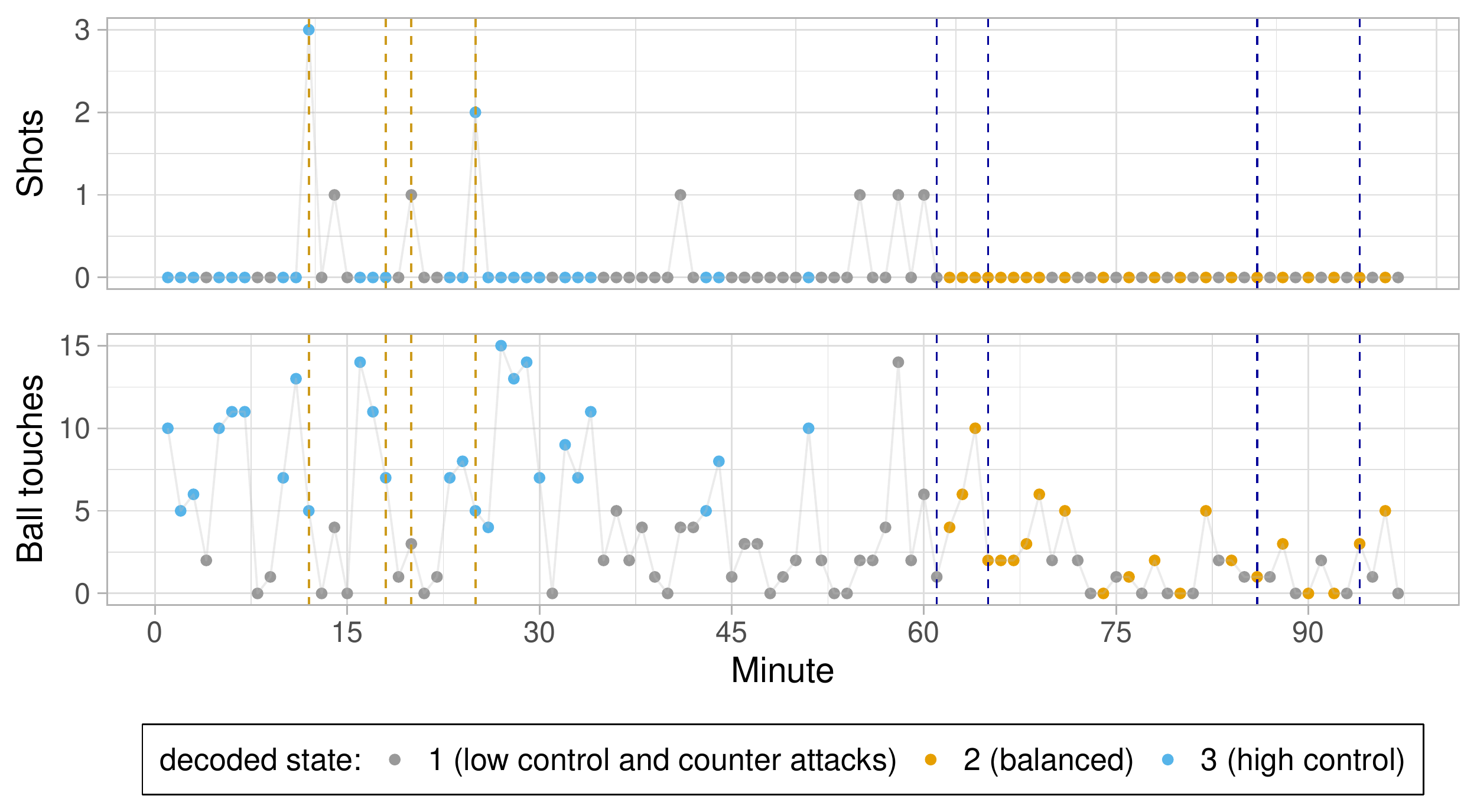}
\caption{Decoded most likely state sequence of the match Borussia Dortmund against Schalke 04 according to the three-state Clayton-copula HMM including covariates. The vertical dashed lines denote goals scored by Borussia Dortmund (yellow lines) and Schalke 04 (blue lines). The goal leading to the intermediate score 2-0 was an own goal by Schalke 04.} 
\label{fig:viterbi}
\end{figure}

To further investigate typical patterns of momentum shifts according to the state process $\{s_{mt}\}$, we calculate the most likely trajectory of the states for match \textit{m}. Specifically, for a given match \textit{m}, we seek
$$ (s_{m1}^*,\ldots,s_{mT_m}^*) = \underset{s_{m1},\ldots,s_{mT_m}}{\operatorname{argmax}} \; \Pr ( s_{m1},\ldots,s_{mT_m} | \mathbf{y}_{m1},\ldots, \mathbf{y}_{mT_m} ), $$
i.e.\ the most likely state sequence, given the observations. 
Maximising this probability is equivalent to finding the optimal of $N^{T_m}$ possible state sequences. This can be achieved at computational cost $\mathcal{O}(T_mN^2)$ using the Viterbi algorithm (see \citealp{zucchini2016hidden}, p.\ 88--92). 
Figure \ref{fig:viterbi} displays the decoded sequences for the match Borussia Dortmund against Schalke 04 which was already shown in Figure \ref{fig:data}. We see that Borussia Dortmund started the match in the high-control state with occasional switches to the low control state with quick counter attacks. According to the decoded state sequence, Borussia Dortmund was in the high-control state for most of the first half, and scored three of their four goals while in that state. After the half-time break, Borussia Dortmund was primarily in the low-control state with quick counter attacks for about 15 minutes, and subsequently alternated between this and the balanced state. In the entire second half, Borussia Dortmund only once was in the high-control state.

At this point it is worth emphasising that our fitted HMM cannot be expected to fully represent all structure and dynamics related to momentum shifts. First, when applied in an unsupervised setting as was done here, then an HMM's model states will generally only be proxies for genuine states \citep{leos2017analysis}. Second, while conceptually appealing and mathematically convenient, it is not necessarily clear that different levels of control and hence momentum shifts are adequately represented by only finitely many states (cf. \citealp{otting2020hot}). Thus, the actual sequence of control levels may of course differ from the decoded sequence as shown in Figure~\ref{fig:viterbi}, and not every inferred state switch refers to a genuine switch in the actual momentum. However, as Borussia Dortmund was occupying the high-control state for most of the first half, but only once in the second half, the decoded state sequence is in agreement with the momentum shift around halftime as suggested by the media.

\begin{table}[ht]
\centering
\caption{Stationary distributions when fixing the score difference at certain levels. Probabilities were calculated for each value of the score difference, with the market value of the opponent and the minute of the match fixed at 200 and 80, corresponding to situations in the final stage of a match against an opposing team of average quality.} 
\label{tab:scorediff_stat}
\scalebox{0.8}{
\begin{tabular}{rrrrrrrrrrrrr}
  \hline
 &   -6 & -5 & -4 & -3 & -2 & -1 & 0 & 1 & 2 & 3 & 4 & 5 \\ 
  \hline
state 1 & 0.073 & 0.100 & 0.134 & 0.175 & 0.222 & 0.280 & 0.523 & 0.732 & 0.705 & 0.642 & 0.560 & 0.475 \\ 
state 2 &  0.391 & 0.364 & 0.334 & 0.301 & 0.267 & 0.234 & 0.206 & 0.175 & 0.147 & 0.122 & 0.098 & 0.076 \\ 
state 3 & 0.535 & 0.535 & 0.532 & 0.524 & 0.511 & 0.486 & 0.271 & 0.094 & 0.148 & 0.236 & 0.342 & 0.450 \\ 
   \hline
\end{tabular}}
\end{table}

\FloatBarrier
\section{Discussion}

There is wide interest in the dynamics of football matches, and specifically in potential momentum shifts, in particular by fans and the media.
From a managerial perspective, it is important to understand the causes of such shifts, and hence also how to potentially exert an influence on the match outcome.
With data sets on in-game summary statistics becoming freely available, we now have the opportunity to statistically investigate the corresponding processes. 
To that end, here we provide a modelling framework --- copula-based multivariate HMMs --- which naturally accommodates potential changes in the dynamics of a match by relating the observed in-game match statistics to latent states. 
A key strength of the proposed approach is that we not only partition a given match into different phases but also allow for the investigation into drivers of how a match unfolds dynamically over time. 
Such in-game modelling could also be useful for bookmakers to obtain more precise estimations of betting odds. For instance, when modelling the time until the next goal during a football match, bookmakers could take into account the latent dynamics of a match as modelled here.

In our exploratory case study, we tested the feasibility of our approach by analysing minute-by-minute data on matches of one particular team, namely Borussia Dortmund.
The underlying states of the fitted model correspond to match phases where Borussia Dortmund exhibits a low level of control with quick counter attacks, to phases where the match is balanced, and to those with high level of control. In addition, the estimated effects of the covariates shed some light on what kind of events may lead to switches between those states. Specifically, we found that Borussia Dortmund has the highest probability of being in the high-control state when having a clear lead or when trailing.

Although the states of the fitted models are tied to different levels of control, it remains unclear to what extent these can be attributed to shifts in the underlying momentum. Inference into the existence of potential momentum shifts is generally challenging given the absence of any formal definition of what constitutes momentum in sports. Without a clear definition, especially the relation between tactical changes and momentum shifts remains unclear --- depending on the definition, it may be necessary to clearly differentiate between these, or alternatively tactical changes may at least need to be taken into account when investigating momentum shifts.
In our case study, some of the reported effects may clearly arise from tactical considerations rather than momentum shifts. For example, for one-goal leads, switching to the low control and quick counter attacks state may of course be a tactical consideration rather than a shift in the underlying momentum. 
The data considered here does not allow us to disentangle these two possible causes, rendering a definitive conclusion whether the switches between the states are momentum shifts or tactical considerations impossible. However, with the states and effects of the covariates considered (cf.\ Figure \ref{fig:coveffects} and Table \ref{tab:scorediff_stat}) being easy to interpret, they still provide interesting insights to dynamics of football matches. 

A clear limitation of the approach as presented here is that we focus on the in-game dynamics of only one of the two teams involved in a match, when in fact it is clear that the dynamics of a match result from the combination of both teams' actions. 
One way to achieve this would be to consider or even construct variables for the state-dependent process which reflect the actions of both teams, e.g.\ both teams' ball touches, or one team's proportion of ball touches in any given minute.
It then seems conceptually appealing to jointly model both teams' underlying latent state corresponding to their exertion of control over the match, which could be achieved using a bivariate Markov chain, resulting in $N^2$ combinations of states (see, e.g., \citealt{sherlock2013coupled, pohle2020primer}). In these model formulations, both teams' underlying state variables are allowed to interact.
To further improve the realism of these models, it would be beneficial to also include tracking data, e.g.\ by considering the distances run per minute as covariate information.

The modelling framework used in the present contribution, i.e.\ copula-based HMMs for modelling football minute-by-minute data, can easily be transferred to other sports for further investigations and possible characteristics of momentum shifts. These sports include, e.g., basketball, where the variables to be modelled comprise, for example, the number of points/shots, the number of rebounds and the number of blocks/steals. More general, sports with two individuals or teams competing against each other and multiple variables measured on a fine-grained scale are best suitable for analysing momentum shifts using the modelling framework provided here.

\newpage

\bibliographystyle{apalike}
\bibliography{refs}

\FloatBarrier
\newpage
\section{Supplementary Material}
\section*{Coefficients in the model for Borussia Dortmund}

\begin{table}[ht]
\centering
\caption{Estimates of the coefficients determining the state transition probabilites as functions of covariates, in the final three-state Clayton copula HMM for the Borussia Dortmund data; 95\% confidence intervals in brackets.}
\label{tab:appendix_coefficients}
\scalebox{0.7}{
\begin{tabular}{rcccccc}
  \hline
 & 1$\rightarrow$2 & 1$\rightarrow$3 & 2$\rightarrow$1 & 2$\rightarrow$3 & 3$\rightarrow$1 & 3$\rightarrow$2 \\ 
  \hline
intercept & -1.447  & -7.749  & -1.918  & -4.922  & -1.474 & -4.111 \\ 
 & [-1.844;\, -1.049] & [-12.14;\, -3.362] & [-2.754;\, -1.082] & [-7.339;\, -2.505] & [-2.147;\, -0.801]  & [-6.430;\, -1.791] \\
  score difference & 0.074 & 1.310 & 0.812 & -4.504 & -0.240 & -0.410 \\ 
   & [-0.207;\, 0.355] & [-0.140;\, 2.760] & [0.197;\, 1.426] & [-7.993;\, -1.015] & [-0.803;\, 0.324]  & [-0.952;\, 0.133] \\
  home & 0.099 & 0.412 & 1.101 & -0.553 & -0.228 & 0.763 \\ 
     & [-0.412;\, 0.610] & [-2.051;\, 2.875] & [0.234;\, 1.968] & [-2.233;\, 1.128] & [-1.315;\, 0.858]  & [-1.064;\, 2.590] \\
  market value & 0.634 & 4.403 & -1.823 & 3.211 & 0.312 & 0.047 \\ 
     & [0.279;\, 0.989] & [1.721;\, 7.086] & [-2.955;\, -0.690] & [1.438;\, 4.983] & [-0.110;\, 0.733]  & [-0.830;\, 0.925] \\
  minute & -0.104 & 6.239 & -1.318 & 4.451 & 0.278 & 2.148 \\ 
     & [-0.443;\, 0.235] & [2.483;\, 9.995] & [-1.876;\, -0.760] & [1.231;\, 7.670] & [-0.225;\, 0.780]  & [0.905;\, 3.391] \\
   \hline
\end{tabular}}
\end{table}

\section*{Additional analysis of Hannover 96 data}

\begin{figure}[!htb]
\centering
\includegraphics[scale=0.5]{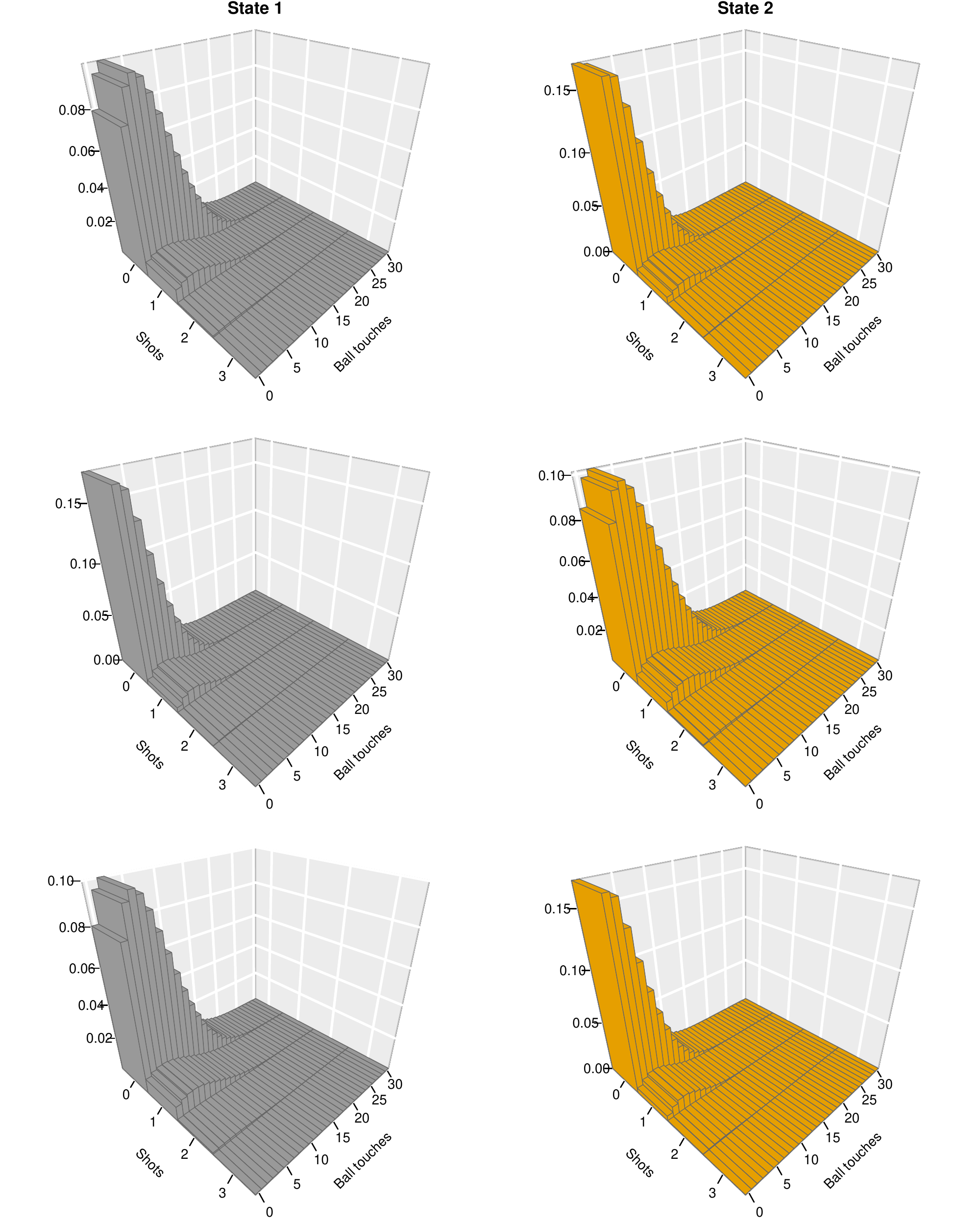}
\caption{Fitted state-dependent distributions for the baseline two-state HMM for Hannover 96. From top to bottom: Frank-, Clayton-, and AMH-copula.} 
\label{fig:baselineH96}
\end{figure}

For the analysis of Hannover 96, we use the same copula-based HMM model formulation as above for Borussia Dortmund. The state-dependent distributions for the fitted baseline model are shown in Figure \ref{fig:baselineH96}. As for Borussia Dortmund, the choice of the copula function considered does not seem to change the shape of the distribution remarkably. Compared to the state-dependent distributions of Borussia Dortmund (see Figure \ref{fig:baselineDO}), Hannover 96 has fewer ball touches and shots on goal, which is intuitively plausible. For all copulas considered, state 1 refers to a high level of control, whereas state 2 can be interpreted as a low level of control. 

\begin{figure}[!htb]
\centering
\includegraphics[scale=0.7]{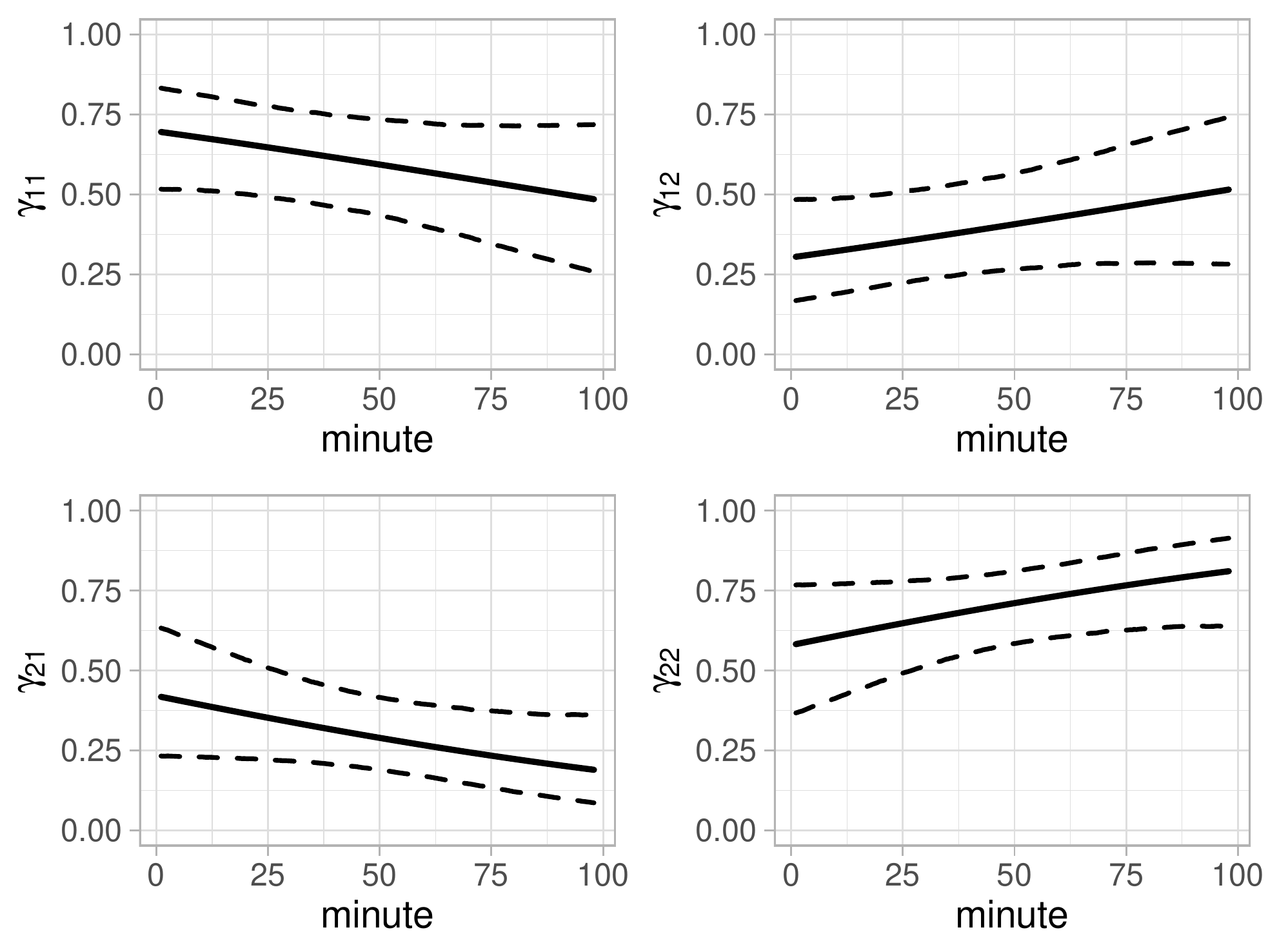}
\caption{Transition probabilities as functions of the covariate minute.} 
\label{fig:coveffects_H96}
\end{figure}

To select a model for Hannover 96, we again compare the AIC and BIC values for different number of states and copulas, which are shown in Table \ref{tab:AICBIC_Hannover}. For the model selected by the BIC, i.e.\
the AMH-copula-based HMM with two states, the transition probabilities as functions of the covariate minute are shown in Figure \ref{fig:coveffects_H96}. As chosen above for Borussia Dortmund, the values for the score difference and the market value of the opponent are fixed at 0 and 200. According to the estimated effects, staying in state 1 (high level of control) becomes less likely at the end of such matches, whereas staying in state 2 (low level of control) becomes more likely. 
The stationary distributions for given values of the score difference are shown in Table \ref{tab:stationary_Hannover}. The values of the minute and the market value of the opponent are again fixed at 80 and 200. We see that the probability for being in state 1 (high-control state) increases if Hannover is trailing. If the score is even or if they are leading, it is more likely that they are in state 2 (low control state) than in state 1, which again is intuitively plausible.

\begin{table}
\caption{AIC and BIC for copula-based HMMs with different numbers of states (Hannover 96).}
\label{tab:AICBIC_Hannover}
\centering
\begin{tabular}{lll:ll:ll}
 & \multicolumn{2}{c}{Frank} & \multicolumn{2}{c}{Clayton}  & \multicolumn{2}{c}{AMH} \\ 
 & AIC & BIC & AIC & BIC & AIC & BIC  \\ \hline
2 states & 18,951 & \textbf{19,030} & 19,024 & 19,103 & 18,949 & \textbf{19,027} \\
3 states & 18,949 & 19,089 & 18,950 & \textbf{19,090} & 18,948 & 19,088 \\
4 states & \textbf{18,888} & 19,101 & 18,911 & 19,123 & 18,920 & 19,132 \\
5 states & 18,891 & 19,789 & \textbf{18,899} & 19,197 & \textbf{18,886} & 19,184 \\
\end{tabular}
\end{table}

\begin{table}[ht]
\centering
\caption{Stationary distributions when fixing the score difference at certain levels. Probabilities were calculated for each value of the score difference, with the market value of the opponent and the minute of the match fixed at 200 and 80, corresponding to situations in the final stage of a match against an opposing team of average quality.}
\label{tab:stationary_Hannover}
\begin{tabular}{rrrrrrrrr}
  \hline
& -4 & -3 & -2 & -1 & 0 & 1 & 2 & 3 \\ 
  \hline
state 1 & 0.638 & 0.642 & 0.626 & 0.539 & 0.320 & 0.111 & 0.028 & 0.006 \\ 
state 2 & 0.362 & 0.358 & 0.374 & 0.461 & 0.680 & 0.889 & 0.972 & 0.994 \\ 
   \hline
\end{tabular}
\end{table}

\end{spacing}
\end{document}